# High coercivity cobalt carbide nanoparticles processed via polyol reaction: A new permanent magnet material

V.G. Harris [1,2], Y. Chen[1], A. Yang[1], S. Yoon[1], Z. Chen[1], Anton Geiler[1], C.N. Chinnasamy[1], L.H. Lewis[3],

C. Vittoria [1,2]

[1]Center for Microwave Magnetic Materials and Integrated Circuits and

[2]Department of Electrical and Computer Engineering, and [3]Department of Chemical Engineering

Northeastern University, Boston, MA 02115 USA

E.E. Carpenter and K. J. Carroll

Chemistry Department, Virginia Commonwealth University, Richmond, VA USA

R. Goswami

SAIC, Washington, DC 20004 USA

M. A. Willard

Code 6355, Naval Research Laboratory, Washington, DC 20375 USA

L. Kurihara

Code 6340, Naval Research Laboratory, Washington, DC 20375 USA

M. Gjoka

Institute for Materials Science, NCSR "Demokritos", Athens, Greece

O. Kalogirou

Department of Physics, Aristotle University of Thessaloniki, Thessaloniki, Greece

**Abstract-** Cobalt carbide nanoparticles were processed using polyol reduction chemistry that offers high product yields in a cost effective single-step process. Particles are shown to be acicular in morphology and typically assembled as clusters with room temperature coercivities greater than 4 kOe and maximum energy products greater than 20 KJ/m$^3$. Consisting of $Co_3C$ and $Co_2C$ phases, the ratio of phase volume, particle size, and particle morphology all play important roles in determining permanent magnet properties. Further, the acicular particle shape provides an enhancement to the coercivity via dipolar anisotropy energy as well as offering potential for particle alignment in nanocomposite cores. While Curie temperatures are near 510K at temperatures approaching 700 K the carbide powders experience an irreversible dissociation to metallic cobalt and carbon thus limiting operational temperatures to near room temperature.

**Introduction**

High performance permanent magnets, i.e. those having appreciable energy products, can be broadly classified into three categories: (i) rare earth intermetallics ($Nd_2Fe_{14}B$ and $Sm_xCo_y$ (where typically x=1 or 2 and y=5 or 17, respectively), AlNiCo (alloys composed primarily of aluminum, nickel and cobalt, with the addition of iron, copper, and sometimes titanium), and ceramic magnets (typically strontium-doped barium hexaferrites). Commercial permanent magnet applications include those for exerting tractive and repelling forces (e.g. magnetic separators, latches, torque drives, bearings, etc.), energy conversion (e.g. magnetos, generators, alternators, eddy current brakes, motors and actuators), directing and shaping particle beams and electromagnetic waves (cathode ray tubes, traveling wave tubes, klystrons, cyclotrons, ion pumps, etc.), and providing magnetic bias fields for a wide range of rf, microwave, and mm-wave devices (isolators, circulators, phase shifters, filters, etc.). The magnets containing rare earth elements provide the highest energy products, $BH_{max}$, but are expensive and prone to corrosion, and as such, pose severe cost limitations and supply chain challenges to commercial industries. Alternatively, AlNiCo and ceramic magnets have substantially lower $BH_{max}$ values but are significantly less expensive and more readily available from many sources. For that reason AlNiCo and ceramic ferrite have defined substantial global permanent magnet market segments. In the case of ceramic magnets the annual revenue generated is second only to NdFeB.

The last major development in viable permanent magnet materials came with the development of NdFeB in the early 1980s, whereas, AlNiCo and ceramic magnets have not experienced significant improvement in permanent magnet properties for several decades. Here, we report a ferromagnetic material based upon nanoscale cobalt carbide particles that provide a rare-earth-free alternative to high performance permanent magnets. The cobalt carbide-based magnets described herein are processed by chemical polyol reduction of metal salts. The precipitate of the reaction need only be rinsed and dried prior to packaging. Packaging may be in the form of isotropic or anisotropic high density compacts, bonded magnets, particle suspensions, etc. The best properties of the carbide particles of the present work include room temperature coercivities greater than 4 kOe and room temperature saturation magnetization up to 60 emu/g. Further, there is a clear dependency upon processing conditions in which the tendency to increase coercivity varies inversely with saturation magnetization. The highest room temperature $BH_{max}$, the primary figure of merit for permanent magnets, is >20 KJ/m$^3$ for free (i.e. not compacted) carbide powders. This value has not yet been optimized for composition and processing but nonetheless compares favorably to both AlNiCo and ceramic magnets as free powders, making these carbide particles a potentially viable and competitive room temperature permanent magnet material.

There has been much attention paid to carbon-containing magnetic materials, which have many potential applications such as in high-density magnetic recording media, high resistivity soft magnetic materials, magnetic toner in xerography, and as contrast agents in high resolution magnetic resonance imaging. In previous work, researchers have focused on cobalt/carbide related materials which include carbon-coated magnetic-metal nanocrystallites,[1] Co-C granular films,[2-5] $M_nC$ (M=Fe, Co, Ni, Cu, n=1-6) nanoclusters[6] and $Co_2C$ films.[7] In those earlier works, the focus was placed on fabrication and application of carbon-related composites. The granular magnetic films, consisting of isolated particles suspended in a nonmagnetic host, were expected to produce low noise hard magnetic media. The so-called core-shell nanoparticles, which constitute another form of nanocomposite, is the core-shell structured nanoparticle; In the 1990s, McHenry et al., reported on the magnetic properties of carbon coated cobalt nanocrystallites.[8] These nanocrystallites where proposed for applications ranging from recording media to emerging biomedical applications in imaging and cancer remediation therapies. Additional research includes theoretical and experimental studies of $M_nC$ (M=Fe, Ni, Co, etc.) clusters,[9] which are cage-like structures of transition metal containing carbon atoms, that demonstrate unusual structural and chemical stabilities; These studies however did not explore, nor report, magnetic properties.

**Chemical Processing and Experimental Procedures**

The chemical synthesis methods employed here to produce size-, shape-, composition- and phase-controlled, highly-coercive cobalt carbide nanoparticles are based upon reduction of metallic salts in a liquid polyol medium that acts as both solvent and reducing agent. The reduction reaction kinetics of the process are enhanced by controlling the type, temperature, and concentration of the polyol medium and by adding appropriate surfactants that limit the oxidation of ions when reduced and regulate the growth of particles as the reaction evolves. The chemical synthesis includes adding 0.1 M - 0.2 M concentration of Co (II) acetate to tetraelthylene glycol. Poly-vinylpyrrolidone (PVP, MW~40,000) was introduced as a capping agent along with NaOH as a catalyst. The solution was allowed to degas in $N_2$ gas (or in some instances Ar gas) for 10-15 minutes prior to the start of the reaction. The solution was then heated to the boiling point of tetraethylene glycol (~573K) for 1-2 h using a distillation apparatus with magnetic stirring. After the completion of the reaction, the solution was cooled to room temperature, magnetically separated several times using an external rare earth magnet, and rinsed repeatedly in methanol. The black precipitate was dried under vacuum at room temperature prior to characterization.

The dried powders were characterized by X-ray diffraction (XRD), transmission electron microscopy, and vibrating sample magnetometry (VSM) for the determination of phase, morphology, and temperature dependent magnetic properties. XRD measurements were performed using a Rigaku-Ultima-III Bragg-Brentano diffractometer employing Cu-Kα radiation (λ=0.15418 nm) in the θ-2θ powder diffraction geometry. Thermomagnetometry was performed using a Lakeshore Cryotonics Inc. Model 7400 VSM for temperatures ranging from room temperature to 1000 K. A Quantum Design physical property measurement system (PPMS) was employed to extend the temperature studies to 10 K. The powders were characterized using a JEOL 2200-FX analytical high-resolution transmission electron microscope with a 200 kV accelerating voltage. Samples for transmission electron microscopy (TEM) were prepared by dispersing a drop of nanoparticl-loaded liquid suspension onto a carbon film supported by copper mesh (400 grid mesh) followed by evaporation of the liquid medium. Fast-Fourier transforms (FFTs) and inverse-fast Fourier transforms (IFFTs) were obtained from the experimental high resolution TEM (HRTEM) images using Digitalmicrograph™ software. Energy dispersive x-ray spectroscopy (EDS) was utilized to determine the composition of the individual powder particle.

**Results & Discussion**

Figure 1 shows a representative θ−2θ x-ray diffraction scan obtained from powders processed using the polyol reduction method described herein. In figure 1, the raw data collected at room temperature from powders that were chemically processed, rinsed, and dried, is depicted with an overlay of data obtained from JCPDS reference powder diffraction files $Co_2C$ (65-1457) and $Co_3C$ (26-0450) in which the intensity and position of each Bragg diffraction peak is represented by a vertical line.

There exist some diffraction features, for example near 67 degrees in 2θ, whose amplitude arises from residual phases that may include different allotropes of carbon and/or unreacted precursors. As yet, residual phases have not been identified; however, these may be related to the apparent surface coating observed in Fig. 2. XRD analysis confirms that $Co_2C$ and $Co_3C$ are the dominant phases present in these nanoparticles.

Figures 2 (a-c) depict high-resolution transmission electron microscopy images. TEM observations show agglomerated particle clusters, about 300-500 nm in diameter (see inset to Fig. 2 (a)), consisting of nanocrystalline Co-carbide particles with acicular or rod-like morphology having an approximate 2:1 aspect ratio. The ferromagnetic nature of these particles is the driving force behind particle agglomeration. Figures 2 (a and b) are TEM images of rod-like Co-carbide crystals. These crystals are surrounded by a thin, 1 to 4 nm graphite-like layer, denoted by arrows in Figs. 2 (a and b). Such a graphitic layer may form during synthesis from the reduction of precursors and surfactants and may act as a barrier that impedes crystal growth. Such a graphitic layer appears in these samples as a layered structure and has been previously reported as an "onion-like" structure in carbide nanoparticles.[10] Figure 2(c) is an HRTEM image of a rod-like Co-carbide nanoparticle with an aspect ratio near 5:1. In order to determine the crystal structure and preferred growth directions, fast Fourier transforms (FFT) were obtained from HRTEM images of individual nanocrystalline particles. Figure 3 (a) is a HRTEM image of a $Co_3C$ nanoparticle with orientation close to the [010] zone axis. The FFT seen in Fig. 3 (b) was obtained from part of the crystal and indexed to the $Co_3C$ phase (space group: Pnma with a=5.03Å, b=6.73Å and c=4.48Å) with additional reflections appearing due to double diffraction. A simulated diffraction pattern of $Co_3C$ along this zone axis is provided for comparison (see Fig. 3(c)). The corresponding inverse fast Fourier transform (IFFT) image (Fig. 3 (d)) shows the lattice spacing of about 5 Å, consistent with a [100] direction along the long axis of the particle. Figure 4 shows a HRTEM image of a $Co_2C$ crystal (space group: Pnnm with a= 4.45Å, b = 4.37Å, and c=2.90Å) close to the [001] zone axis. The FFT (Fig. 4 (b)) from a portion of the crystal shows a near-square pattern indicative of this zone in which the lattice parameters a and b are nearly equal. In this zone, the (100) and (010) reflections are present due to double diffraction (Fig. 4 (b) in red). The corresponding IFFT image shows the lattice spacing of (100) and (010) is ~4.4 Å. Such analyses confirm the following: (i) that $Co_3C$ and $Co_2C$ coexist in the carbide nanoparticles, (ii) the carbide nanoparticles have an acicular morphology with the aspect ratio varying between phases and preparation conditions from 2:1 to 7:1, and (iii) the crystallites are surrounded by a thin graphite-like layer.

Table I presents the phase volume ratios and lattice parameters of each phase determined by Rietveld reduction analyses of the XRD data for several samples. In addition to the these data derived from XRD analyses, similar data from selected area electron diffraction (SAED) as well as values reported in the literature from bulk standards are presented[11, 12] The XRD and SAED determined lattice parameters are consistent with bulk values within the uncertainty of the measurements and analyses.

Figure 5 is a room temperature hysteresis loop curve of one cobalt carbide nanoparticle sample. For this sample the room temperature magnetization under an applied field of ~17 kOe is 73 emu/g with a coercivity of 3.1 kOe. For the purposes of this study we report the magnetization at 17 kOe as the saturation magnetization ($M_s$) although it is clear that complete saturation was not attained. This sample has a room temperature $BH_{max}$ of 20.7 KJ/m$^3$. Figure 6 presents the room temperature saturation magnetization and coercivity data for many particle samples collected over the course of this study. There exists a great variation of properties coinciding with a broad range of chemical process parameters. However, it is clear from figure 6 that there is a trade-off in magnetic properties, that is, the greater the saturation magnetization the lower the coercivity. We have found that these magnetic properties can be controlled by careful selection of process parameters and coincide with variations in the $Co_2C:Co_3C$ volume fraction and relative particle size and morphology of each phase. The error bars presented on figure 6 data points represent the uncertainty in the measurement of saturation magnetization due to the ambiguity in the volume and mass of the particle sample. All magnetization values have been corrected for the presence of the nonmagnetic graphitic surface layer. The correction involved the calculation of the surface layer volume based upon the thickness measured in HRTEM images and assuming a rectangular cross section leading to the renormalization of the magnetic moment.

Figure 7 is a plot illustrating the interrelationship between saturation magnetization and coercivity to the volume fraction of $Co_2C$ to $Co_3C$ measured by x-ray diffraction (see Table I). One sees that as the relative fraction of $Co_2C$ increases the magnetization reduces while concomitantly the coercivity increases. This result suggests the role of each carbide phase. For example, one may conclude that $Co_3C$ possesses a relatively large saturation magnetization while $Co_2C$ is responsible for the large coercivity. It does not however, indicate the fundamental origin of the large room temperature coercivity measured in these samples. Since the particles are clearly acicular in morphology one can conclude that dipolar or shape anisotropy is responsible for some fraction of the coercivity. Further, the atomic structure in these phases deviates from cubic symmetry and therefore a second source of anisotropy is expected to be of a magnetocrystalline nature. Other sources of loss may be related to exchange between particles. Such interparticle exchange, including that of $Co_2C$-$Co_2C$, $Co_2C$-$Co_3C$, and $Co_3C$-$Co_3C$, may provide yet other significant contribution to anisotropy, and subsequently coercivity, in these nanoparticle carbide systems.

Thermomagnetic properties of a representative carbide powder sample are presented in figures 8 and 9. Figure 8 illustrates the temperature response of magnetization for a sample heated from 10 K to 900 K. Magnetization data was collected as a function of temperature under the application of 0.5 kOe and 10 kOe fields. The data of figure 8, collected under the application of a 10 kOe field, began at 10 K and approached a Curie temperature of ~510 K. The solid curve is a fit to a molecular field approximation. At temperatures approaching 700 K a dramatic increase in magnetization is measured. The thermal cycle reveals an irreversible transformation. The magnetization and high Curie temperature of the sample heated above 700 K is consistent with metallic cobalt. We conjecture that during this vacuum heat treatment the carbide disassociates to metallic cobalt and free carbon. Having a Curie temperature near 510 K and lacking high temperature chemical and structural stability these materials may be presently limited to room temperature permanent magnet applications.

It has been established that these carbide nanoparticles exist in two phases, namely $Co_2C$ and $Co_3C$. The room temperature hysteresis loop of figure 5 illustrates a continuous variation of magnetization through remanence; behavior that is consistent with a single magnetic phase or the exchange coupling of the two carbide phases. Figure 9 contains both 300 K and the 10 K hysteresis loops of a representative sample and clarifies this assertion. At 10 K, a knee is observed near remanence indicating the decoupling of hard and soft phases. From the trends seen in figure 7, the soft phase is likely $Co_3C$. These results infer that the $Co_2C$ and $Co_3C$ phase are exchange coupled at room temperature. If the exchange is of a particle-particle nature, or as an admixture of the two phases, is as yet unknown.

**Summary and Conclusions**

In summary, the magnetic and structural properties of cobalt carbide nanoparticles processed using a single-step polyol reduction reaction were presented. Particles are shown to be acicular in morphology, typically assembled as clusters, with room temperature coercivities in some samples greater than 4 kOe and maximum energy products greater than 20 KJ/m$^3$. Consisting of $Co_3C$ and $Co_2C$ phases, the ratio of phase volume and particle morphology determine permanent magnet properties. In figure 10 the energy products of cobalt carbide nanoparticles are compared with free powders of AlNiCo and ceramic magnets of the Ba/Sr hexaferrite type. Curie temperatures near 510 K indicate suitability for room temperature permanent magnet applications, however at temperatures approaching 700 K the carbide particles disassociate irreversibly to carbon and metallic cobalt. The large room temperature coercivities originate from dipolar and magnetocrystalline anisotropies of carbide phases but may be additionally affected by interparticle exchange. The discovery of large room temperature energy product in nanoparticle cobalt carbides provides a new and viable rare earth-free permanent magnet material.

**Acknowledgements**

This research was funded by the Office of Naval Research under contract N000140910590.

**References**

[1] Wang Z H, Choi C J, Kim B K, Kim J C and Zhang Z D, 2003 *Carbon* **41** 1751–1758

[2] Lee Y H, Huang Y S, Min J F, Wu G M and Horn L, 2007 *J Magnetism and Magnetic Materials* **310** 913–915

[3] Konno T J, Shoji K, Sumiyama K and Suzuki K, 1999 *Journal of Magnetism and Magnetic Materials* **195** 9-18

[4] Wang H, Wong S P, Cheung W Y, Ke N, Lau W F, Chiah M F and Zhang X X, 2001 *Materials Science and Engineering C* **16** 147–151

[5] Zeng F H and Zhang X, 2007 *Journal of Magnetism and Magnetic Materials* **309** 160–168

[6] Black S J, Morley C P, Owen A E, Elsegood M R J, 2004 *J. Organometallic Chemistry* **689** 2103–2113

[7] Premkumar P A, Turchanin A and Bahlawane N 2007 *Chem. Mater.* **19,** 6206–6211

[8] McHenry M E, Majetich S A, Artman J O, DeGraef M and Staley S W, 1994 *Phys. Rev. B* **49**, 11358

[9] Zhang Z X, Cao B B and Duan H M, 2008 *J. Molecular Structure: THEOCHEM* **863** 22–27

[10] Huh S H and Nakajima A, 2006 *J. Appl. Phys.* 99, 064302

[11] Clarke J, 1951 *Chem. Ind. (London)*, **46** 1004

[12] General Electric Company, Corporate R. and D., Schenectady, New York, USA. Private communication (1974)

**Table Captions**

Table I: Structural properties determined by X-ray diffraction and electron diffraction measurements

**Figure Captions**

Figure 1. A representative θ−2θ x-ray diffraction scan obtained from powders processed using the polyol reduction reaction. Vertical lines corresponding to the position and amplitude of diffraction peaks from JCPDS reference powder diffraction files $Co_2C$ (65-1457) and $Co_3C$ (26-0450) are also presented.

Figures 2 (a-c). High-resolution transmission electron microscopy images of a representative cobalt carbide sample. The insert to (a) shows an agglomerated particle cluster about 300-500 nm in diameter. Panels (a) and (b) are TEM images of rod-like Co-carbide crystals surrounded by a thin 1 to 4 nm graphite-like layer (denoted by arrows). Figure 2(c) is an HRTEM image of a rod-like Co-carbide nanoparticle of aspect ratio near 5:1.

Figures 3 (a-c). HRTEM image of a $Co_3C$ nanoparticle with orientation close to the [010] zone axis. The FFT (Fig. 3 (b)) was indexed to the $Co_3C$ phase (space group: Pnma with a=5.03Å, b=6.73Å and c=4.48Å) with additional reflections appearing due to double diffraction. A simulated diffraction pattern of $Co_3C$ along this zone axis is provided for comparison in (c). The corresponding IFFT image (d) shows the lattice spacing of about 5 Å, consistent with a [100] direction along the long axis of the particle.

Figures 4 (a-c). HRTEM image of a $Co_2C$ crystal (space group: Pnnm with a= 4.45Å, b = 4.37Å, and c=2.90Å) close to the [001] zone axis. The FFT (b) from a portion of the crystal shows a near-square pattern indicative of this zone in which the lattice parameters, a and b, are nearly equal. The (100) and (010) reflections are present due to double diffraction (b) in red. The IFFT image indicates the lattice spacing of (100) and (010) is ~4.4 Å.

Figure 5. Room temperature hysteresis loop of representative sample having Ms of 73 emu/g and an Hc of 3.1 kOe. The $BH_{max}$ is 20.7 kJ/m$^3$.

Figure 6. Room temperature saturation magnetization ($M_s$) and coercive field ($H_c$) data for many particle samples collected over the course of this study. Saturation magnetization values correspond to moments measured under the application of 17 kOe.

Figure 7. A plot of magnetic properties versus the phase volume ratio illustrating the interrelationship between saturation magnetization and coercivity to the volume fraction of $Co_2C$ to $Co_3C$ measured by x-ray diffraction.

Figure 8. Temperature response of magnetization for a representative sample heated from 10 K to over 900 K. The magnetization was collected under the application of a 10 kOe field. The solid curve is a fit to a molecular field theory approximation with $T_C$ of 510K. At temperatures approaching 700 K an irreversible transformation occurs. The high magnetization and Curie temperature of the sample heated above 700 K is consistent with metallic cobalt.

Figure 9. 300 K and the 10 K hysteresis loops of a representative sample. At 10K a knee is observed near remanence indicating the decoupling of hard and soft phases. Results infer that $Co_2C$ and $Co_3C$ phases are exchange coupled at room temperature but not at 10K.

Figure 10. Energy products of cobalt carbide nanoparticle powders compared with powders of alnico and ceramic permanent magnet systems.

**Table I: Structural properties determined by X-ray diffraction and electron diffraction measurements**

| Sample No. | Volume Ratio $Co_2C : Co_3C$ | Lattice Parameters of $Co_2C$ (Angstrom) | | | Lattice Parameters of $Co_3C$ (Angstrom) | | |
|---|---|---|---|---|---|---|---|
| | | a | b | c | a | b | c |
| Bulk standards | | 4.371 | 4.446 | 2.897 | 4.444 | 4.993 | 6.707 |
| SAED | | 4.37 | 4.45 | 2.90 | 4.48 | 5.03 | 6.73 |
| 4-4 (2) | 1.89:1 | 4.361 | 4.446 | 2.888 | 4.448 | 5.005 | 6.718 |
| 4-4 (3) | 1.06:1 | 4.362 | 4.444 | 2.891 | 4.450 | 5.002 | 6.712 |
| 4-4 (4) | 0.93:1 | 4.365 | 4.443 | 2.894 | 4.454 | 4.998 | 6.714 |
| 4-4 (5) | 0.99:1 | 4.364 | 4.443 | 2.900 | 4.443 | 5.005 | 6.710 |
| 4-4 (8) | 1.46:1 | 4.361 | 4.444 | 2.890 | 4.454 | 5.004 | 6.707 |

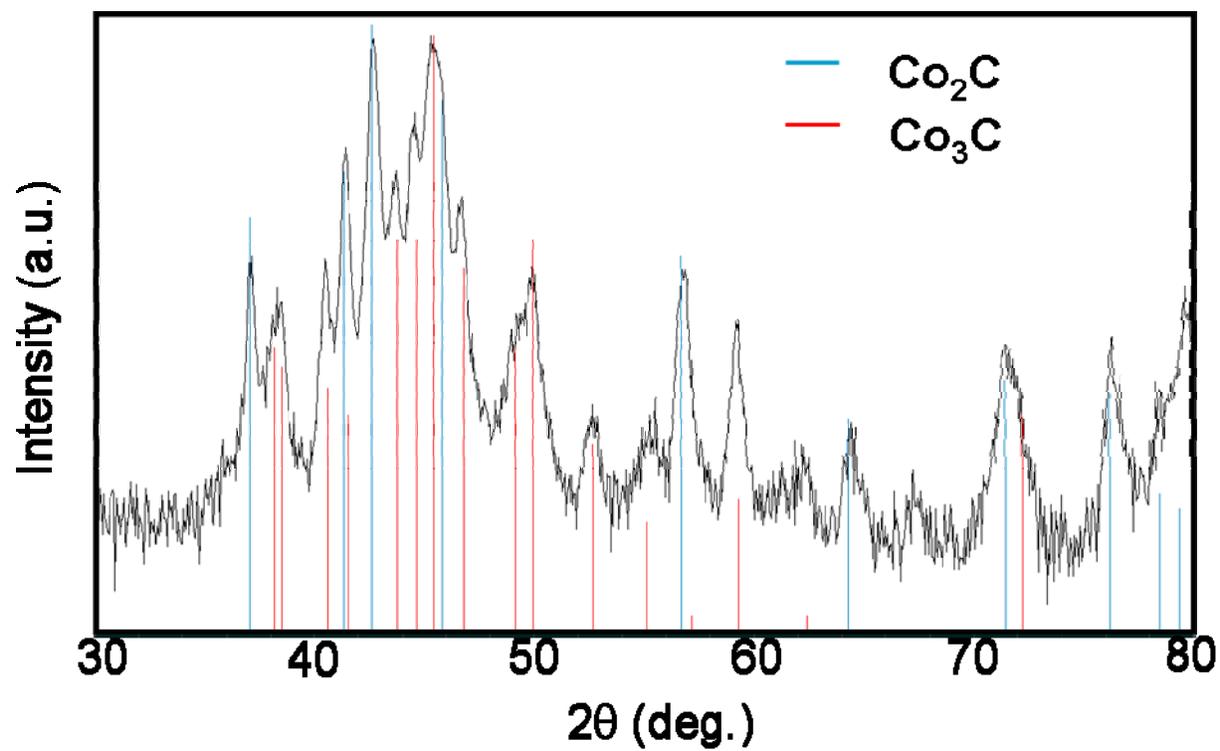

**Figure 1**

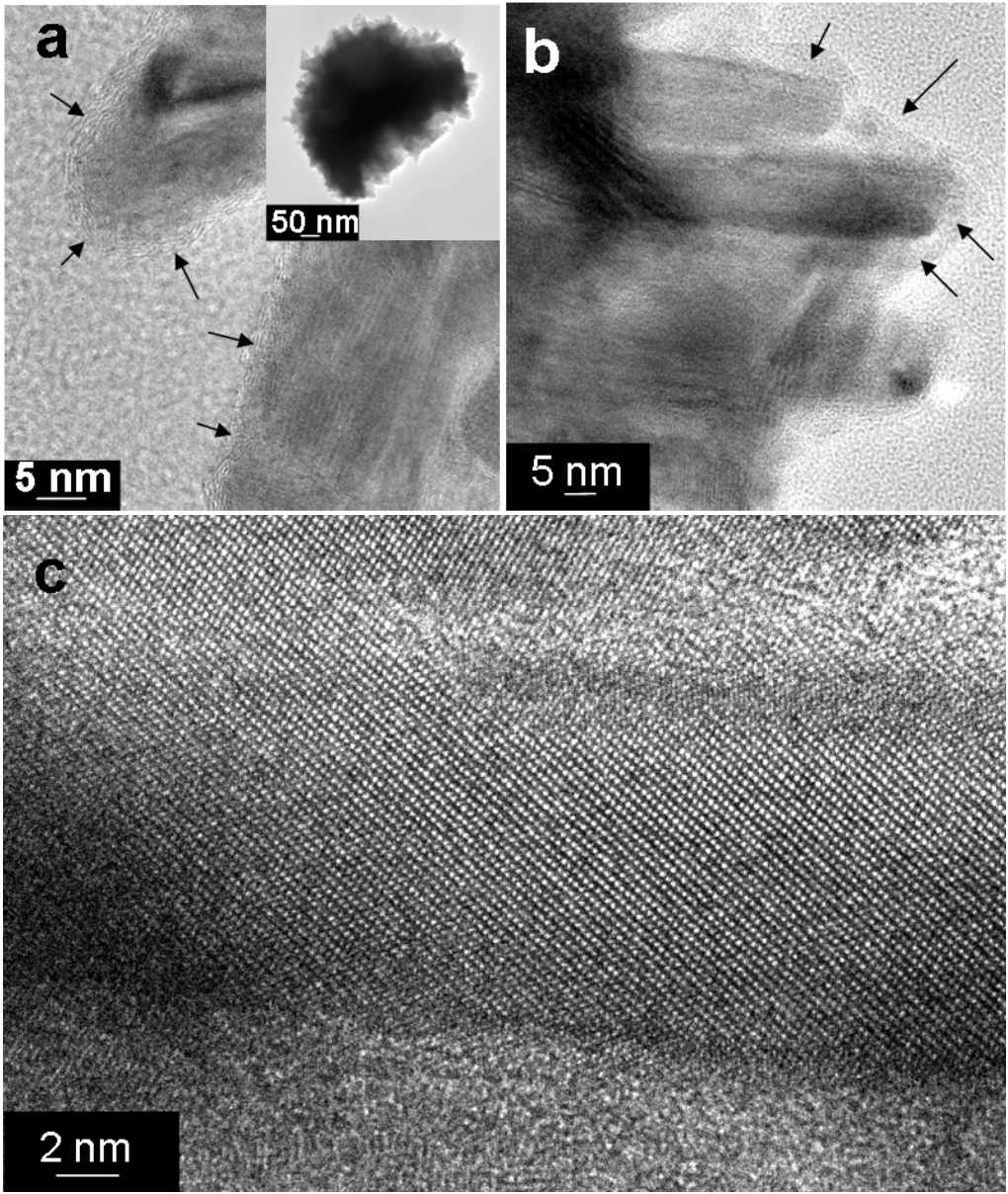

**Figure 2**

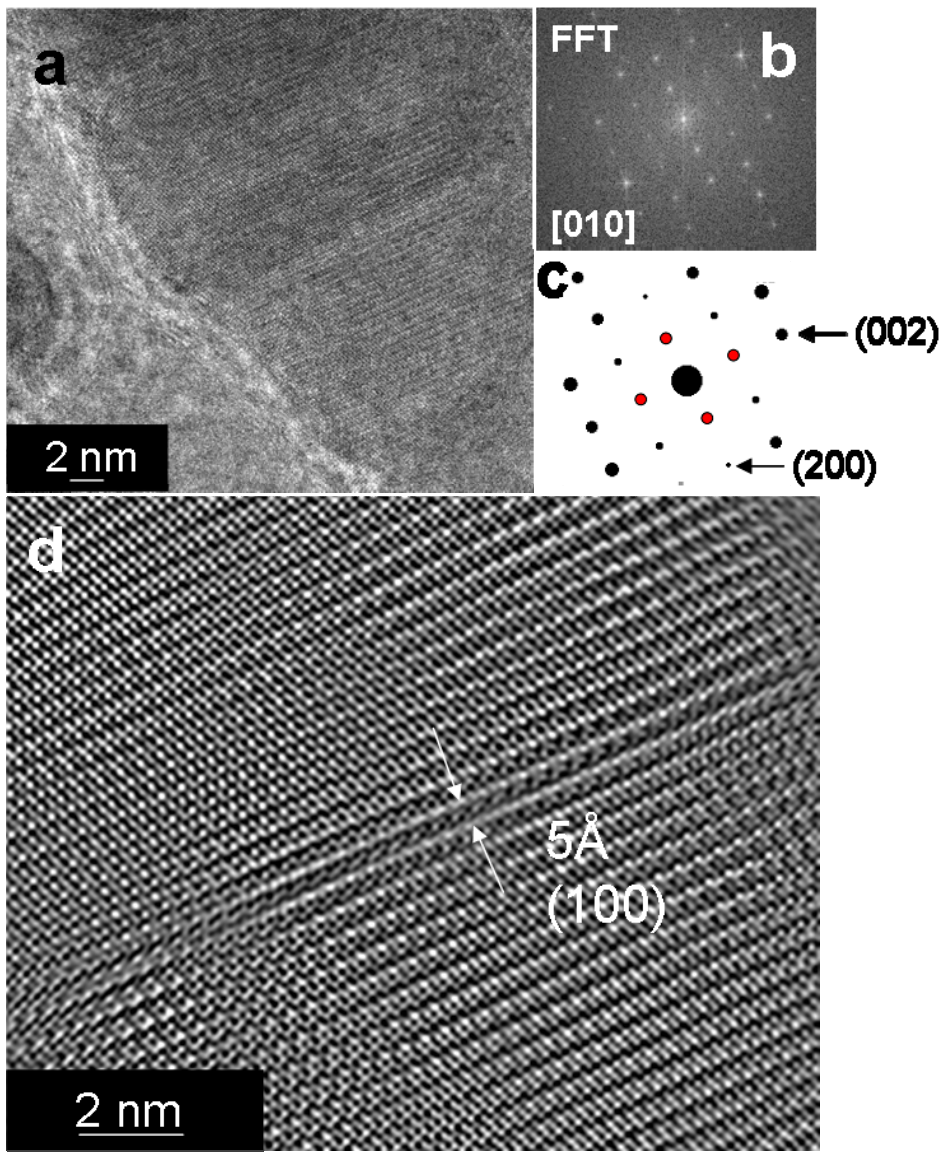

**Figure 3**

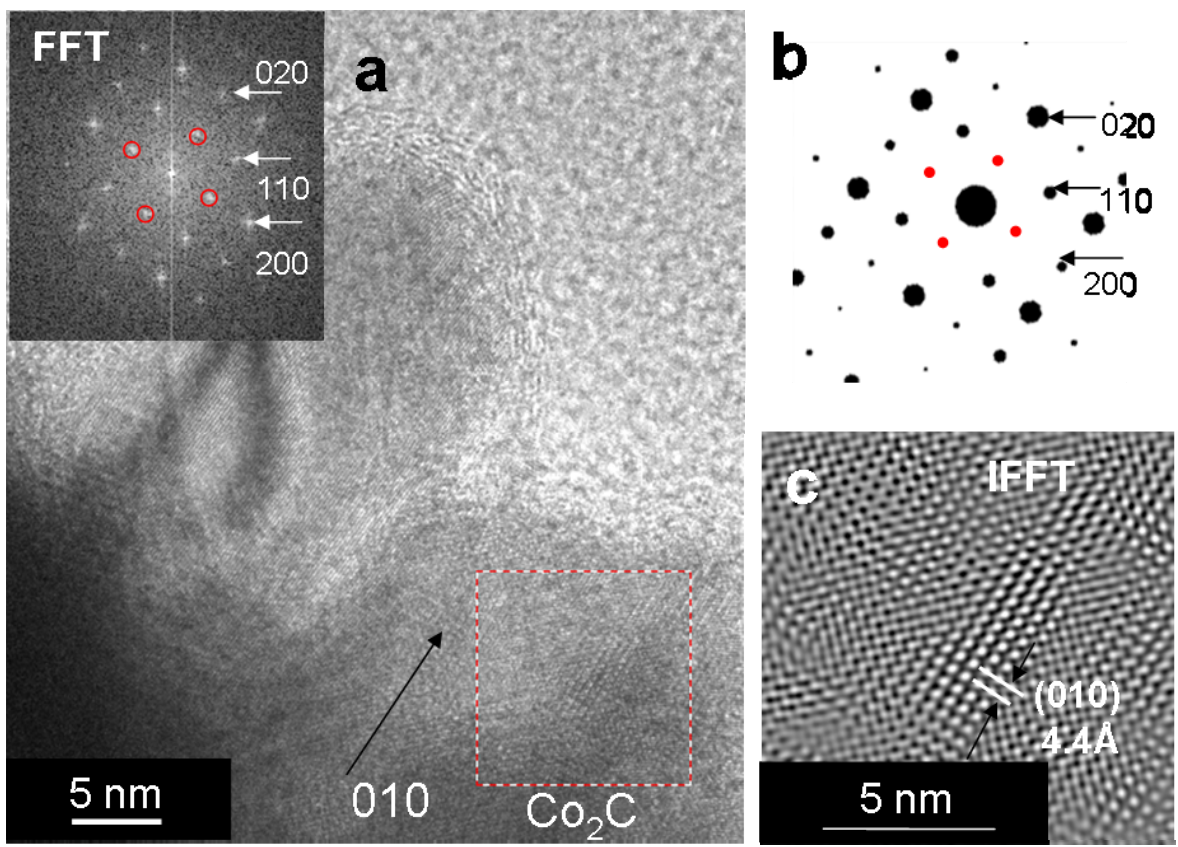

Figure 4

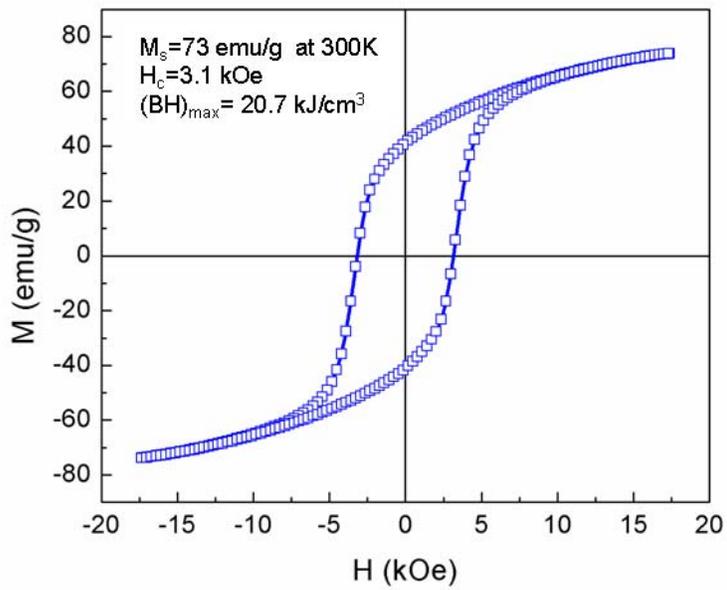

Figure 5

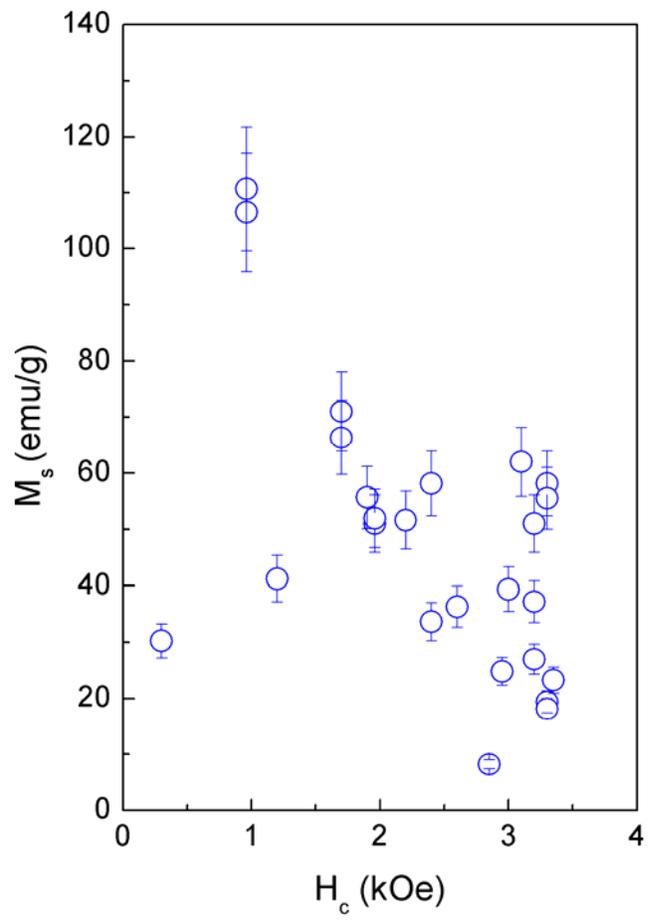

Figure 6

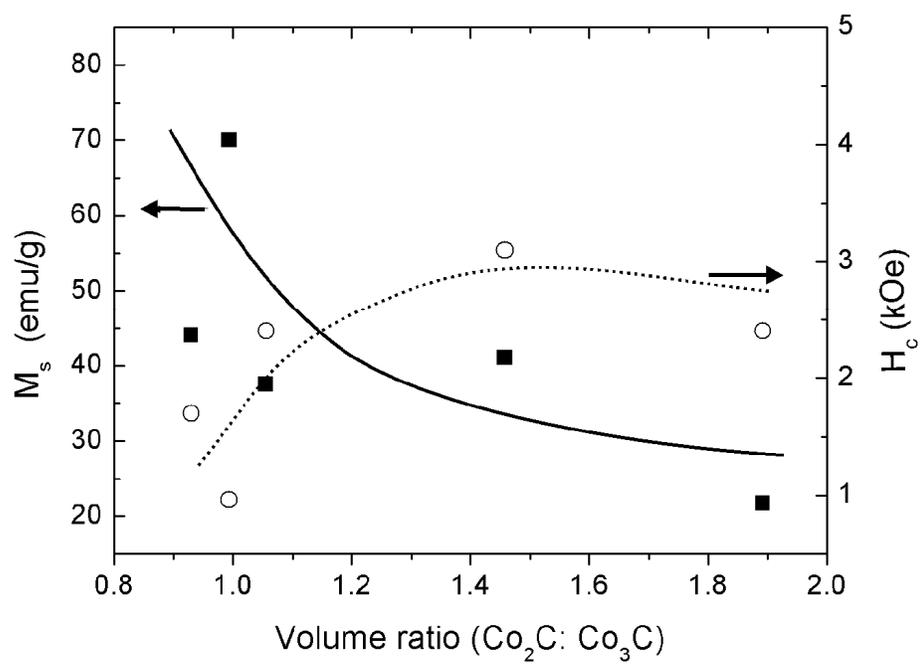

Figure 7

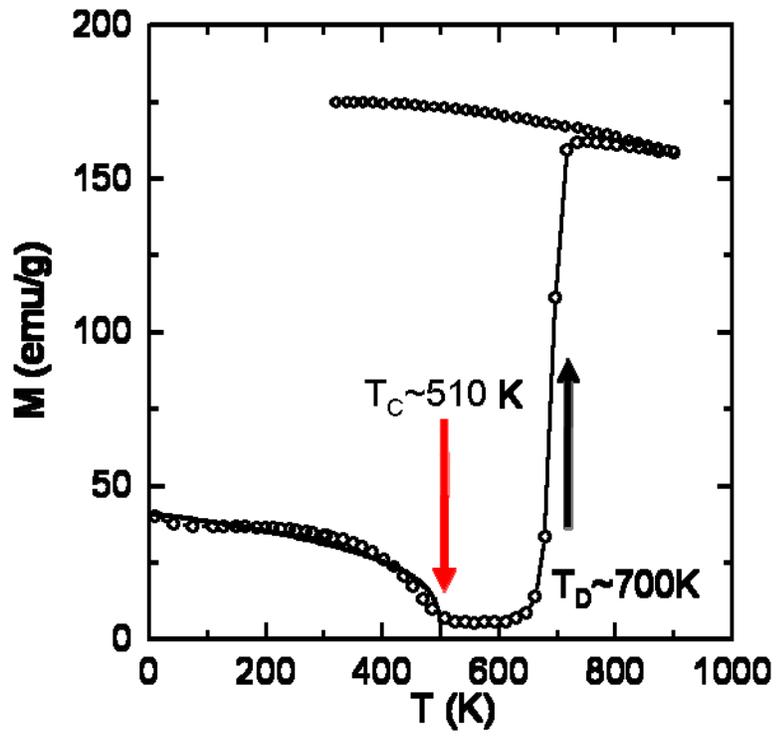

Figure 8

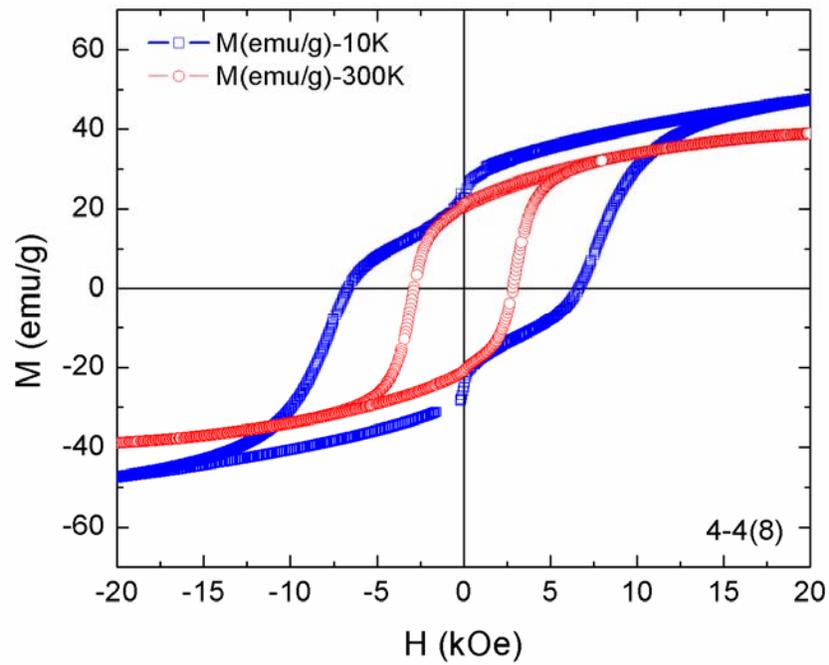

Figure 9

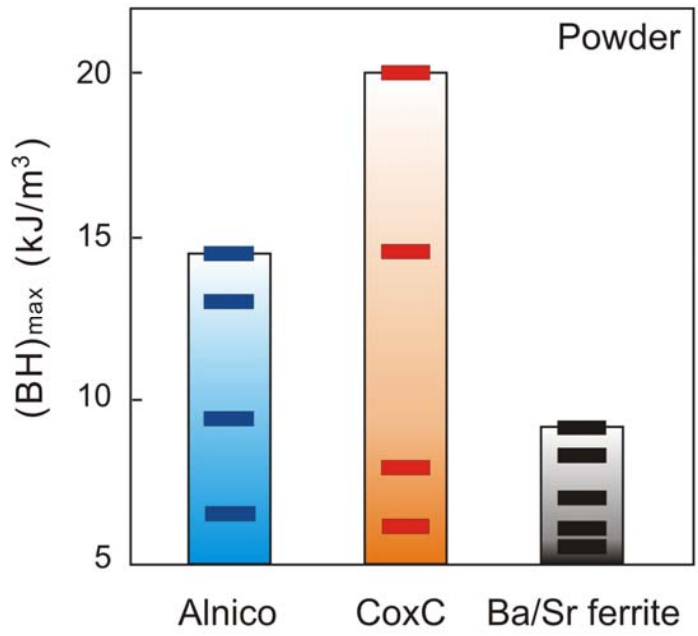

Figure 10